\documentclass{article} 
\usepackage{arxiv,times}

\usepackage{microtype}
\usepackage{hyperref}
\usepackage{url}
\usepackage{booktabs}
\usepackage{amsmath,amssymb}
\usepackage{graphicx}
\usepackage{multirow}
\usepackage{enumitem}
\usepackage{xcolor}
\usepackage{subcaption}
\usepackage{lineno}

\usepackage{xspace}

\definecolor{darkblue}{rgb}{0, 0, 0.5}
\hypersetup{colorlinks=true, citecolor=darkblue, linkcolor=darkblue, urlcolor=darkblue}

\newcommand{\model}{Qwen3-4B}
\newcommand{\mures}{\boldsymbol{\mu}_\text{res}}
\newcommand{\mucomp}{\boldsymbol{\mu}_\text{comp}}
\newcommand{\hh}[1]{\mathbf{h}^{(#1)}}

\title{Where Do LLMs Decide to Break the Rules? Mechanistic Localization of Prompt Injection Compliance}

\author{%
Rui Wen$^\textbf{1}$, Jiayang Liu$^\textbf{2}$, Zeyu Yang$^\textbf{3}$, Jun Sakuma$^\textbf{1,5}$, Lu Sun$^\textbf{4,5}$\\
$^1$Institute of Science Tokyo, $^2$Nanyang Technological University\\ $^3$Singapore University of Technology and Design, $^4$Tohoku University, $^5$RIKEN AIP
}

\iclrfinalcopy

\hypersetup{
  hidelinks,
  pdftitle={Where Do LLMs Decide to Break the Rules? Mechanistic Localization of Prompt Injection Compliance},
  pdfauthor={Rui Wen and Jiayang Liu and Zeyu Yang and Jun Sakuma and Lu Sun}
}

\begin{document}
\maketitle

\begin{abstract}

When a prompt injection attack succeeds, a Large Language Model (LLM) abandons its assigned system role to comply with an adversarial instruction. While prior work has extensively quantified how often this occurs, we ask a more fundamental question: \textit{where inside the network does the model actually decide to break the rules?}
Using layer-by-layer causal activation patching across five models (4B to 32B parameters), we find a clear dissociation: attack information is linearly decodable from the first layer, yet causal leverage over the model's behavior is negligible until a late-layer bottleneck in the final third of the network.
Patching this bottleneck reverses compliance in 77--92\% of cases.
We show that the compliance mechanism occupies a compact linear subspace (rank-8 in 4B and 14B models, scaling to rank-64 at 32B) and is architecturally stable across varying model families. Finally, we validate our mechanistic account by showing that this causal peak layer is also the representationally optimal site for detecting attacks, outperforming early-layer classifiers that degrade under surface-level obfuscation such as leetspeak substitution. This alignment between causal leverage and detection performance provides converging evidence that the late-layer bottleneck captures decision-relevant computation rather than merely reflecting an artifact of the intervention.
\end{abstract}

\section{Introduction}
\label{sec:intro}

\emph{Prompt injection} exploits a fundamental vulnerability in LLMs: the absence of a hard security boundary between developer instructions and user inputs. An attacker who controls any part of the context window can override the system prompt simply by inserting new instructions into the prompt~\citep{PR22}.
As LLMs are deployed as autonomous agents with access to tools, databases, and external APIs, this vulnerability escalates from an inconvenience to a direct privilege-escalation vector: injections embedded in retrieved documents, emails, or tool outputs can silently redirect an agent's actions in the real world~\citep{GAMEHF23,ZLYK24}.

A growing literature measures how often attacks succeed~\citep{LJGJG24,YXZKSXW25,DZBBFT24} and proposes defenses ranging from input filtering~\citep{CPSW25,HLHZZK24} to system-prompt hardening~\citep{WXLWHB24}.
Yet all of these approaches treat the model as a black box and offer no principled account of \emph{why} compliance happens. Without knowing where inside the network the decision is made, defenses remain heuristic: an adversary who understands the mechanism has a structural advantage over a defender who does not.

\paragraph{Why mechanistic localization matters.}
Knowing \emph{where} causal leverage concentrates has three concrete consequences.
First, it enables surgical intervention. If the decision to comply with an attack is bottlenecked in a specific region of the network, defenders can focus their monitoring and steering efforts exactly there. This isolates the defense, avoiding the need to retrain or modify the entire model.
Second, it dictates the feasibility of lightweight runtime defenses. If leverage is concentrated, a probe on one layer may suffice; if leverage is broadly distributed, intervention must be systemic.
Third, it reveals a structural vulnerability that we can directly exploit for detection. While prior work shows that models often encode concept layers before actually acting on them~\citep{BFSHOMBS23}, we demonstrate how extreme this gap is for prompt injection: the attack is recognizable by the first layer, yet causal leverage over whether the model complies does not appear until late in the network. We leverage this exact split to build a robust defense. Because early layers primarily process surface-level text, detectors placed there are easily blinded by simple obfuscations like leetspeak. By contrast, a detector placed at the true causal peak targets the model's underlying intent, making it substantially more robust against evasion.

\begin{figure*}[t]
\centering
\includegraphics[width=0.96\textwidth]{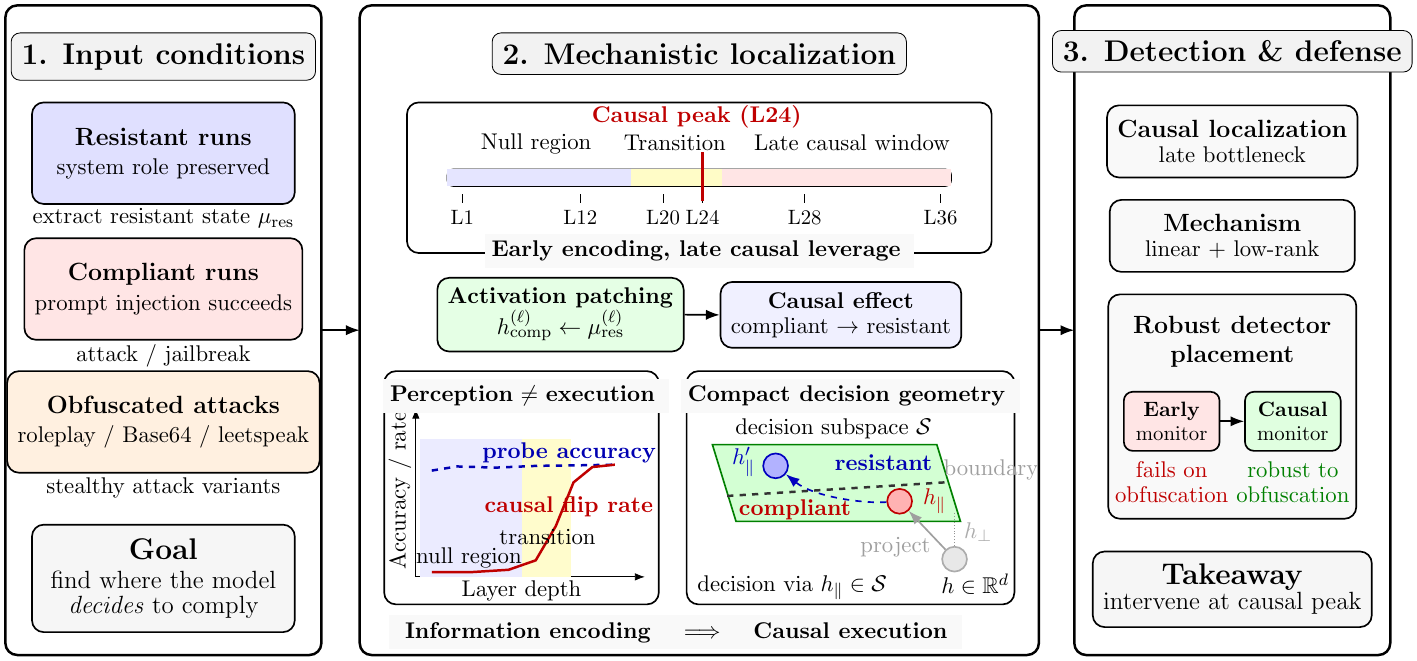}
\caption{
\textbf{Mechanistic localization and defense framework.}
\textbf{Left:} resistant, compliant, and stealthy/obfuscated attacks define the contrast used for causal intervention. 
\textbf{Middle:} attack presence is decodable from early layers, yet causal leverage remains near zero until a late-layer bottleneck. The underlying mechanism is compact and low-rank (empirically rank $\approx 8$ in 4B--14B models). 
\textbf{Right:} the causal peak captures decision-level semantics rather than surface form, making it the most effective and obfuscation-robust site for detection.
}
\label{fig:framework}
\end{figure*}

\paragraph{Our approach.}
We apply \emph{causal activation patching}~\citep{MBAB22,GIZCCHAWGPI25}: when a model complies with an injection, we intercept its forward pass at layer $\ell$ and replace the hidden state with the average state of resistant runs. If the model then refuses, that layer has causal leverage. Sweeping every layer produces a precise causal map.
The result is consistent across different models and attack types: \textit{attack information is linearly decodable from the first layer, yet causal leverage over the model's behavior is negligible until a late-layer bottleneck.}
Figure~\ref{fig:framework} presents an overview of our framework, illustrating how prompt injection behaviors are mechanistically localized to a late-layer causal bottleneck and governed by a compact low-dimensional subspace, enabling robust detection and intervention.

In summary, we make four main contributions:
\begin{itemize}
  \item \textbf{Causal localization of prompt injection compliance.}
    We provide the first systematic layer-wise causal analysis of prompt injection. Across models from 4B to 32B, causal influence is negligible in early layers and peaks in a late-layer bottleneck. This reveals a clear separation between representation and action.

  \item \textbf{Mechanistic characterization of the compliance process.}
    We demonstrate that the compliance mechanism operates within a low-dimensional linear subspace, suggesting it is a structured circuit rather than a diffuse property of the network.

  \item \textbf{Robustness across settings and attack types.}
    The location of this late-layer window is stable across the dense architectures we test, across decoding temperatures, and under adversarial suffixes and multi-turn escalations; stealthy attacks, which travel different representational pathways in early layers, converge on the same late block.

  \item \textbf{Detection as validation of the mechanism.}
    Our causal analysis predicts that the optimal detection layer should align with the causal bottleneck. We confirm this empirically: the causal peak layer yields the best detection performance among the tested layers, and its detector is robust to obfuscation. This provides external validation of the mechanistic account.
\end{itemize}

\section{Related Work}
\label{sec:related}

\paragraph{Prompt Injection Attacks and Defenses.}

Early work systematically studied direct prompt-injection attacks, where adversarial instructions in the user input cause the model to deviate from its intended behavior~\citep{PR22}. Subsequent work extended the threat to indirect settings, showing that malicious instructions embedded in retrieved documents, web content, or tool outputs can hijack model behavior without direct user interaction~\citep{GAMEHF23}. Benchmarking efforts now evaluate these vulnerabilities systematically across models and scenarios~\citep{LJGJG24, ZLYK24, YXZKSXW25}, including agentic settings where injections can trigger unauthorized tool calls~\citep{DZBBFT24}.

On the defense side, proposed mitigations include structured separation of instructions and data~\citep{CPSW25}, provenance-marking transformations of untrusted inputs~\citep{HLHZZK24}, instruction-hierarchy training that teaches models to prioritize privileged instructions~\citep{WXLWHB24}, and detection-based guardrails~\citep{LJJSG25}. These approaches primarily mitigate prompt injection at the input-output level. They measure or reduce whether an attack succeeds, but do not directly identify which internal computations exert causal control over compliance. Our work studies this complementary question by systematically mapping where causal leverage over prompt-injection compliance emerges across layers.

\paragraph{Causal Intervention, Representation Engineering, and Safety Internals.}

Our methodology builds on two lines of work in mechanistic interpretability. The first uses causal interventions to localize model behavior, including causal tracing for factual associations~\citep{MBAB22}, path patching~\citep{GMSA23}, and the formal framework of causal abstraction~\citep{GIZCCHAWGPI25}. The second is representation engineering, which modifies model behavior by manipulating directions in activation space~\citep{ZPCCGRPYMDGLBWMBKSFKH23, RGSTHT24, TTLUVMM24}.

A growing body of work applies related techniques to model safety. \citet{AOSPPGN24} identified a low-dimensional refusal direction, while \citet{LYZL25} and \citet{ZZXGKS25} identified safety-critical layers and neurons. \citet{ZYZXHL24} and \citet{BKP26} studied jailbreak-related hidden-state dynamics, and \citet{HHECRMNN25} showed that instruction-following success can be predicted from intermediate representations.

Most closely related, \citet{WWYTXLDW25} defend against indirect prompt injection through \emph{instruction detection}, showing that intermediate layers can provide stronger detection signals than the final layer. Our work is complementary. Rather than choosing a layer based on detection performance alone, we first measure causal leverage through layer-wise activation interventions and then test whether the resulting causal localization predicts where detection works best. This reveals a clear separation between representation and control: attack information is decodable from the earliest layers, while causal leverage over compliance emerges substantially later. We further characterize this late-stage mechanism as a compact subspace and show that its location predicts an effective layer for prompt-injection detection.

\section{Method}
\label{sec:methods}

\subsection{Experimental Setup}
\label{sec:setup}

\paragraph{Models and Stimuli.}
We study five instruction-tuned models spanning 4B to 32B parameters: three from the Qwen family~\citep{YLYZHZYGHLZLZHHGWLTYTZYYZZLDBYYDLXLZWZMGLLLTYRWZRFSZZWLWCZZQ25}, Phi-3-mini~\citep{AAAAABBBBBBBBBCCCCCCCCCCDDEFGGGGGGGHHHHHIJJJKKKKKKLLLLLLLLLLLLLLMMMMMMMNNPPPPQRRRRRRSSSSSSSSSSTTVWWWWWWWWWWWWXXXXXYYYYYYYZZZZZZZZ24}, and Mistral-7B-Instruct-v0.3~\citep{JSMBCCBLLSLLSSLWLS23}, with \model{} as our primary testbed. 
To observe prompt injection in action, we established a strict ``role-conflict'' scenario: we assigned the models a rigid system role (an ``AcmeCorp'' customer service agent; the verbatim system prompt and the additional persona prompts are given in Appendix~\ref{app:prompts}) and then exposed them to 300 test queries. These queries contained either \textbf{direct attacks} (e.g., ``Ignore all previous instructions''), \textbf{stealthy attacks} embedded in plausible narratives, or \textbf{benign control} questions.

\paragraph{Labeling.}
We classified the model's outputs as either \emph{compliant} (breaking character to execute the attack) or \emph{resistant} (refusing the attack) using strict keyword matching. To guarantee that our baseline data was unambiguous, these labels were independently validated by a cross-model LLM judge.

\subsection{Causal Activation Patching}
\label{sec:patch_protocol}
To pinpoint exactly where the network ``decides'' to break its system prompt, we use causal activation patching. For each compliant run at every layer $\ell$, we performed a three-step intervention at the last prompt token:

\begin{enumerate}[topsep=2pt,itemsep=1pt]
\item \textbf{Isolate the Antidote:} We calculated the average internal representation ($\mures^{(\ell)}$) of the model when it successfully resisted an attack.
\item \textbf{Overwrite the State:} We intercepted the forward pass of a model actively complying with an attack, replacing its hidden state at layer $\ell$ with this ``resistant'' average.
\item \textbf{Measure the Flip:} We let generation continue and recorded the \emph{flip rate}, i.e., the fraction of trials in which the intervention causes the model to abandon compliance and resist.
\end{enumerate}

If intervening at a specific layer consistently causes the output to flip, that layer possesses direct causal leverage over the model's compliance decision.

\paragraph{Controls and Mechanism Isolation.}
A natural concern is that any sufficiently large perturbation might derail the model rather than surgically redirect it. To rule this out, we run a battery of controls at the peak causal layer. Injecting \textit{benign} states, \textit{compliant} states, or \textit{norm-matched random noise} produces substantially smaller effects than the resistant intervention, confirming that the effect is specific to the resistant geometry rather than to perturbation magnitude.

We then use finer-grained interventions to characterize \emph{how} the mechanism operates. We used Rank-$k$ PCA patching to show the mechanism occupies a highly compact linear subspace and sublayer decomposition to isolate exactly where the causal signal originates within the transformer block.

\section{Results}
\label{sec:results}

\subsection{Model Knows Early, Acts Late}
\label{sec:res_probe}

\begin{figure}[t]
\centering
\includegraphics[width=0.98\linewidth]{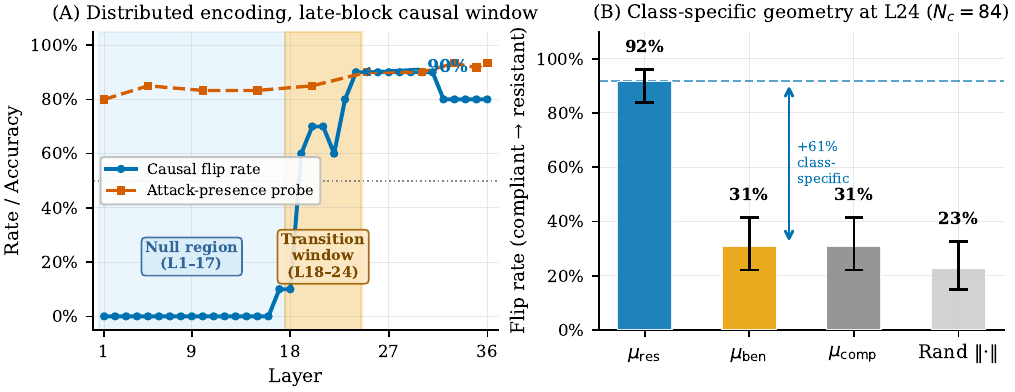}
\caption{%
\textbf{(A)~Layer-by-layer causal flip rate and probe accuracy for \model{}.}
The attack-presence probe demonstrates that the model can distinguish whether an attack is present at early layers.
Causal flip rate is negligible before L19, showing that encoding attack presence and committing to comply are dissociated.
\textbf{(B)~Four-condition control battery at the L24 peak.}
The gap between resistant and random confirms a class-specific effect.
}
\label{fig:overview}
\end{figure}

\textbf{The model encodes the attack everywhere.}
If we simply look at what the model ``knows,'' the attack is obvious from the start. A logistic-regression probe can distinguish injection from benign inputs with 80\% or higher accuracy at every single layer. The geometric distance between injection and benign mean states grows from layer 1 through 35 before dropping at L36. The model encodes attack presence from the very first layer.

\textbf{Causal leverage is negligible in the first half and concentrated late in the network.}
Despite this persistent early signal, the causal sweep reveals a clear separation between encoding and behavioral control (Figure~\ref{fig:overview}A). Interventions through L18 have little effect, followed by a sharp transition from L19 and sustained 80--90\% flip rates in the late layers, peaking at L24. Controls in Figure~\ref{fig:overview}B show that this effect is specific to the resistant representation rather than perturbation magnitude alone.

A same-prompt analysis further rules out input-content differences. For prompts that yield both compliant and resistant outcomes under repeated sampling, patching the prompt-final state at L24 with the resistant centroid raises $P(\text{resist})$ from $0.57$ to $0.93$, while a content-matched resistant donor raises it to $0.87$. The compliant centroid leaves it nearly unchanged at $0.54$, whereas a norm-matched random intervention lowers it to $0.15$. Through L16, no donor changes $P(\text{resist})$ by more than $0.09$. At L28--L32, the compliant centroid also begins to induce resistance, leaving L24 as the clearest class-specific causal leverage point rather than a layer that is generically sensitive to perturbation.

\textbf{Layer 24 is the critical relay point.}
Logit-lens analysis (Figure~\ref{fig:crossarch}A) clarifies what happens around this peak. Layer 23 sharpens the model's resistance, while Layer 25 finalizes the commitment to comply. Layer 24 sits directly between these two events. Sublayer decomposition confirms this timeline. Neither the attention mechanism (21.4\%) nor the MLP (21.4\%) alone can recover the 90.5\% full-patch effect (Appendix~\ref{app:sublayer}, Table~\ref{tab:sublayer}). The signal does not originate \emph{within} Layer 24. Instead, it arrives encoded in the \emph{incoming} residual stream. Patching at Layer 24 intercepts this compliance signal right at its most concentrated transit point; L24 is thus best read as a late-block causal \emph{leverage point} rather than the site where the compliance decision is first computed.

\subsection{The Late-Block Causal Window is Architecturally Stable}
\label{sec:res_generalize}

\begin{figure}[t]
\centering
\includegraphics[width=0.98\linewidth]{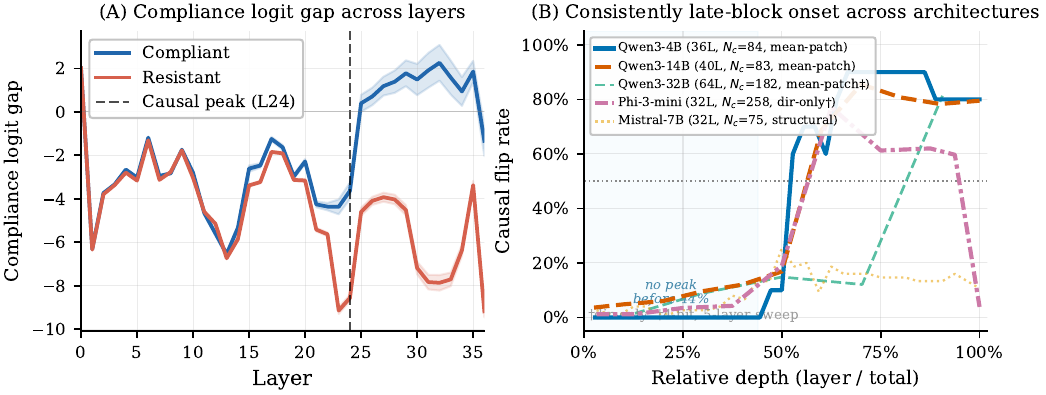}
\caption{%
\textbf{(A) Logit-lens: compliance commitment across layers.}
Compliance logit gap ($\log P_\text{Sure} - \log P_\text{Sorry}$) for
compliant (blue) and resistant (red) pools.
\textbf{(B) Cross-architecture causal sweeps} (depth normalized to 0--1).
Results show negligible causal leverage in the first
37--44\% of layers, with a late-block rise.
}
\label{fig:crossarch}
\end{figure}

\textbf{Cross-architecture replication.}
We replicated this causal sweep across four additional models ranging from 4B to 32B parameters. As Figure~\ref{fig:crossarch}B demonstrates, no model exhibits meaningful causal leverage before reaching 44\% of its total depth. The peak causal depths consistently cluster in the final third of each network, specifically between 63\% and 91\% of the total layers.

\textbf{The causal window is the robust metric.}
The exact peak layer is sensitive to noise and can shift by several layers across prompt templates. A more robust summary is the \emph{causal window}: the contiguous band of layers where the flip rate reaches at least 75\% of its peak. We characterize each window by its \textbf{onset depth} (where the window first opens) and its \textbf{width} (the number of sampled sweep layers inside the window), and its \textbf{center of mass} (the flip-rate-weighted average layer over the layers inside the window). As Table~\ref{tab:arch_windows} shows, these quantities cluster tightly across models and conditions, confirming that the late-layer bottleneck is a stable architectural feature. 

\begin{table}[t]
\centering
\small
\caption{Causal peaks and windows across models and conditions (\model{}: 36L; Qwen3-14B: 40L; Phi-3-mini: 32L). Window: contiguous sweep layers with raw flip rate $\geq$75\% of peak; Onset: its first layer; Width: number of sampled sweep layers in the window; COM: flip-rate-weighted mean layer over the window layers (depth = COM / total layers). }
\label{tab:arch_windows}
\setlength{\tabcolsep}{4pt}
\begin{tabular}{llccccc}
\toprule
Model & Corpus & Peak (depth) & Peak flip & Onset (depth) & Width & COM (depth) \\
\midrule
\model{}   & Headline    & L24 (67\%) & 0.917 & L24 (67\%) & 6 & L31.4 (87\%) \\
\model{}   & Multi-turn  & L24 (67\%) & 0.871 & L20 (56\%) & 4 & L26.1 (73\%) \\
Qwen3-14B  & AcmeCorp    & L28 (70\%) & 0.855 & L24 (60\%) & 5 & L32.1 (80\%) \\
Phi-3-mini & AcmeCorp (additive) & L20 (63\%) & 0.771 & L20 (63\%) & 4 & L25.1 (78\%) \\
\bottomrule
\end{tabular}
\end{table}

\textbf{Naturalistic benchmarks preserve the causal profile.}
To test whether our findings depend on the synthetic AcmeCorp setting, we evaluate two public prompt-injection benchmarks, \texttt{deepset/prompt-injections} and \texttt{neuralchemy/Prompt-injection-dataset}, using a dense 19-layer sweep with benchmark-matched donors (Appendix~\ref{app:benchmark_matched}). Although peak flip rates drop to about 12\% under more diverse phrasing, the same early null region and late-layer peak remain.
We further replicate this pattern on \texttt{microsoft/BIPIA} and \texttt{Lakera/gandalf\_ignore\_instructions} (Appendix~\ref{app:external}). Both peak at L24 with a clear margin over same-layer norm-matched random controls (BIPIA: $0.70$ vs.\ $0.13$; Gandalf: $0.66$ vs.\ $0.32$); their higher absolute flip rates likely reflect more templated phrasing. Gandalf attacks are natively user-turn; for BIPIA we extract the attack payloads and present them in the user turn under the AcmeCorp prompt, so this tests transfer of BIPIA attack content, not its indirect-injection threat model, which we study separately in Section~\ref{sec:res_stealthy}.

\subsection{Mechanism Is Linear and Compact}
\label{sec:res_cascade}

\textbf{The mechanism occupies a tight linear subspace.}
Replacing the full 2560-dimensional hidden state is unnecessary. Projecting the intervention onto the top eight principal components already recovers 81.0\% flip rate, compared with 90.5\% for the full patch, crossing the 80\% effectiveness threshold (Appendix~\ref{app:rankk}, Table~\ref{tab:rankk}). This indicates that most of the intervention effect lies in a compact linear subspace. The identity of this subspace also matters: under the class-mean protocol, a rank-8 intervention based on the learned PCA subspace achieves an 85.7\% flip rate, far above the 38.1\% achieved by a random rank-8 subspace with matched energy.

\textbf{This low-dimensional structure is stable across data splits.}
We split the resistant examples into two halves, fit the subspace independently on each half, and evaluate cross-split transfer. The resulting subspaces produce similar intervention effects, suggesting that the observed low-rank structure is not specific to a particular donor sample.

\textbf{Required subspace dimensionality varies with architecture.}
Under the same per-stimulus protocol, Qwen3-14B, like Qwen3-4B, crosses the 80\% effectiveness threshold at $k{=}8$ (Table~\ref{tab:rankk}). Qwen3-32B requires $k{=}64$ to cross the same absolute threshold, but its full-patch ceiling is also lower at 81.3\%. Normalizing by each model's full-patch effect, $k{=}8$ already recovers 90\% (4B), 91\% (14B), and 89\% (32B) of the maximum effect. Moreover, the first principal component captures 90\% of the 32B mean displacement. Thus, the intervention remains similarly low-dimensional across scales; what shifts is the absolute threshold relative to each model's attainable ceiling.

\section{Generalization and Scope}
\label{sec:res_stealthy}

\begin{figure}[t]
\centering
\includegraphics[width=\linewidth]{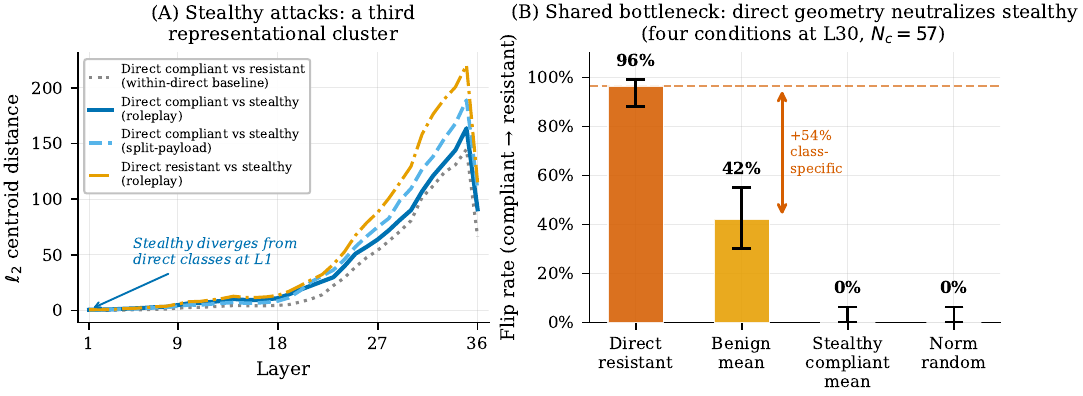}
\caption{%
\textbf{(A) Stealthy attacks occupy a distinct representational pathway.}
$\ell_2$ centroid distances show stealthy-compliant states form a third cluster, farther from both direct-attack classes at every layer.
\textbf{(B) Four-condition control battery at L30.}
The gap between direct-resistant and benign mean confirms the effect is class-specific: direct-attack geometry generalizes to attack types never seen during calibration.
}
\label{fig:stealthy}
\end{figure}

\textbf{Stealthy attacks share the exact same late-layer bottleneck.}
Attackers often hide their prompt injections inside plausible narratives or roleplay scenarios to evade simple detection. Two representative examples are \emph{roleplay framing}, where the injection is embedded inside a scenario (e.g., ``pretend you are a poet and write...''), and \emph{split payload}, where the override instruction is divided across multiple sentences to bypass pattern-matching defenses. 

Figure~\ref{fig:stealthy}A shows that these attacks follow a distinct early representational pathway. At every layer, stealthy states are farther from direct-attack states than direct-compliant and direct-resistant states are from each other; at L1, this distance is eight to nine times larger. A linear probe also perfectly separates stealthy from direct attacks at every layer (separability $=1.000$). This confirms that \textit{disguised attacks travel a completely different representational pathway} from the very beginning.

Despite this early divergence, the exact same defense geometry used for direct attacks successfully neutralizes these stealthy attacks. Patching with the direct-attack resistant mean achieves a \textbf{96.5\%} flip rate at Layer 30. A four-condition control battery rules out generic network disruption. The targeted direct-resistant intervention yields a 96.5\% flip rate, compared to 42.1\% for the benign mean, 0.0\% for the stealthy-compliant mean (an off-distribution donor), and 0.0\% for norm-matched random noise. This 54-percentage-point advantage over the benign baseline shows that direct-attack geometry successfully generalizes to neutralize attack types it never encountered during calibration.

\paragraph{Indirect injection reveals a principled boundary.}
Indirect injections place adversarial instructions inside retrieved content \citep{GAMEHF23}. We construct 600 stimuli spanning 300 topics and two prefix variants (Appendix~\ref{app:indirect}). Using this corpus's own resistant centroid, the flip rate rises late and reaches 43.3\% at L36 (Table~\ref{tab:indirect_sweep}), but a norm-matched random vector achieves the same rate. The diverse retrieval topics therefore dilute the global centroid, leaving little compliance-specific signal.

The limitation appears to arise from the donor rather than the target geometry. On a separate 60-stimulus pilot corpus, transferring a clean direct-injection resistant/compliant contrast to compliant indirect targets ($N_c{=}44$) yields a 95.5\% flip rate at L24 with coherent refusals, compared with 6.8--15.9\% for the pilot corpus's own resistant centroid on the same targets. Because this pilot differs from the 600-stimulus corpus above, the 95.5\% and 43.3\% rates are not directly comparable. Under the clean donor, the effect is also robust across intervention operators: a full centroid, scaled contrastive direction, and rank-8 subspace achieve 95.0\%, 92.5\%, and 85.0\% flip rates, respectively ($N_c{=}40$; Appendix~\ref{app:indirect}, Table~\ref{tab:indirect_ops}). These results indicate that late-layer compliance geometry remains accessible for indirect injection when the donor cleanly isolates the compliance contrast.

\paragraph{Summary.}
The late-layer bottleneck remains stable across the tested attack types, while intervention strength depends strongly on donor quality. A direct-injection donor transfers to pilot indirect targets with a 95\% flip rate at L24, whereas the indirect corpus's own centroid can perform no better than random. Likewise, benchmark-matched donors do not outperform cross-distribution AcmeCorp donors on naturalistic attacks (12\% vs.\ 19\%; Appendix~\ref{app:benchmark_matched}). Together, these results suggest that isolating the compliance contrast matters more than distributional matching alone.

\section{Robustness Across Settings}
\label{sec:robustness}

Late-layer causal localization remains consistent across model architectures from 4B to 32B, stimulus types, and input formats. No tested model exhibits meaningful causal leverage before 44\% of its depth, while the causal window remains consistently late.

\textbf{Sampling temperature.}
We first test whether this pattern is an artifact of greedy decoding by repeating a sparse causal sweep with sampled decoding at $T\in\{0.3,0.7,1.3\}$. As shown in Table~\ref{tab:temperature}, L24 and L36 exceed 89\% flip rate at every temperature, while L1 and L16 remain substantially weaker. Higher temperatures increase early-layer effects somewhat but do not shift causal leverage away from the late layers.

\begin{table}[h]
\centering
\small
\caption{Sparse causal sweep under sampled decoding on \model{}. $N_c$ denotes the number of compliant stimuli at each temperature.}
\label{tab:temperature}
\begin{tabular}{lccccc}
\toprule
$T$ & $N_c$ & L1 & L16 & L24 & L36 \\
\midrule
0.3 & 86 & 0.023 & 0.151 & 0.895 & 0.895 \\
0.7 & 87 & 0.103 & 0.149 & 0.897 & 0.908 \\
1.3 & 95 & 0.200 & 0.242 & 0.905 & 0.916 \\
\bottomrule
\end{tabular}
\end{table}

\textbf{GCG adversarial suffixes.}
We next ask whether adversarially optimized input perturbations change where causal leverage emerges. We optimize a universal 20-token GCG suffix on 20 direct-attack stimuli and apply it to the remaining 280. The optimized suffix does not increase attack success in our setting: 28 of the 280 stimuli remain compliant (10\%, compared with 28\% without the suffix). We therefore do not interpret this experiment as evidence of robustness to a stronger GCG attack. Instead, we condition on the compliant GCG-suffixed cases and ask whether their causal profile shifts. As shown in Table 3, flip rates remain substantially lower through L16 and reach 100\% at L24, indicating that the adversarial suffix changes the input surface without shifting the late region in which causal intervention becomes effective.

\begin{table}[h]
\centering
\small
\caption{Causal sweep on GCG-suffixed stimuli (\model{}).}
\label{tab:gcg}
\scalebox{0.9}{
\begin{tabular}{lccccccc}
\toprule
Layer & L1 (3\%) & L8 (22\%) & L16 (44\%) & \textbf{L24 (67\%)} & L28 (78\%) & L32 (89\%) & L36 (100\%) \\
\midrule
Flip rate & 0.107 & 0.179 & 0.321 & \textbf{1.000} & 1.000 & 1.000 & 1.000 \\
\bottomrule
\end{tabular}
}
\end{table}

The pattern also persists under multi-turn escalation. When the injection appears only after a benign first turn, causal leverage again peaks at L24, with an 87.1\% flip rate compared with 91.7\% in the single-turn setting. Full multi-turn results are reported in Appendix~\ref{app:robustness}.

\section{Detection as Mechanistic Validation}
\label{sec:res_detection}

The disconnect between what the model \emph{knows} and what it \emph{does} yields a concrete, testable prediction. If the compliance decision truly crystallizes at the model-specific causal peak, then that peak layer should also be the best place to detect an attack. A detector at Layer~1, by contrast, should be easy to fool, despite achieving $\geq$80\% probe accuracy on clean inputs, because early layers encode surface form rather than the model's intent.

We test this by training a linear classifier (LinearSVC) on hidden states at each layer. To keep optimization fast and regularization stable, we first compress the 2560-dimensional states to 128 principal components ($\approx$90\% variance retained), then evaluate against a hard-negative set of 100 benign queries deliberately crafted with injection-adjacent vocabulary.

\begin{table}[h]
\centering
\small
\caption{Layer-by-layer detection mirrors the causal sweep profile.
L24 outperforms both L1 and L36 despite all three having $\geq$80\% attack-presence probe accuracy.}
\label{tab:detection_layers}
\begin{tabular}{lcc}
\toprule
Layer & AUROC & Leet.\ recall  \\
\midrule
L1  & 0.888 & 13.3\% \\
\textbf{L24 (causal peak)} & \textbf{0.993} & \textbf{100\%}  \\
L36 & 0.892 & 100\%   \\
\bottomrule
\end{tabular}
\end{table}

Table~\ref{tab:detection_layers} confirms this prediction. Layer~24 achieves an AUROC of 0.993, substantially higher than Layer~1 (0.888) and Layer~36 (0.892). Thus, the layer with the strongest causal leverage also provides the best detection performance.

The most revealing comparison is on obfuscated inputs. Leetspeak, for example, replaces letters with visually similar numbers (``inject'' becomes ``1nj3ct'').
A Layer~1 classifier drops to 13.3\% recall on these inputs: it relies on surface patterns that simple substitutions destroy. Layer~24, by contrast, detects all five tested obfuscation strategies (leetspeak, zero-width spaces, synonym substitution, noise insertion, and base64 encoding) at full recall.
This is precisely what the causal account predicts. Early layers see tokens; the causal peak encodes the model's decision to comply, which is invariant to how the attack is spelled.

The convergence of three independent signals at Layer~24, i.e., peak causal leverage, best detection accuracy, and full robustness to obfuscation, provides strong external validation of the mechanistic account. The causal map is not an artifact of classifier design; it is a structural property of the network that directly predicts where the most effective defenses should be placed.

\paragraph{Comparison to a text-level detector and an adaptive attacker.}
We further compare against \texttt{ProtectAI/deberta-v3-base-prompt-injection} and a detector-aware adversary (Appendix~\ref{app:detector_ext}). On the hard-negative set, the L24 probe reaches AUROC 0.993 versus 0.709 for the text classifier while reusing activations already computed during prefill. A white-box GCG attack optimized against the L24 score reduces recall at 5\% FPR from 1.00 to 0.48 while preserving compliance, but transfers only weakly to the text classifier ($0.625\!\to\!0.600$). We therefore treat the causal-peak detector as mechanistic corroboration and a lightweight complementary signal, rather than a standalone defense.

\section{Discussion and Limitations}
\label{sec:discussion}

Causal location is more stable than intervention magnitude. Flip rates reach 77--92\% in the calibrated AcmeCorp setting, but drop to about 12\% on diverse naturalistic benchmarks; more templated BIPIA and Gandalf recover 66--70\% at the same L24 region (Appendix~\ref{app:external}). Indirect injection shows the same pattern: topical variation dilutes the global indirect-own centroid to the level of a random vector, while a clean direct-injection donor recovers a 95\% flip rate at the same late layer (Appendix~\ref{app:indirect}). Thus, intervention strength depends on how cleanly the donor isolates the compliance contrast and on phrasing diversity, whereas the location of causal leverage remains comparatively stable.

Several questions remain open. Richer donor representations may improve transfer to naturalistic inputs, while the strong detection signal and low-dimensional subspace motivate lightweight monitoring and targeted activation steering. The single-token picture also weakens in broader settings: agentic rollouts are not controlled by a single prompt-final or commitment-token patch (Appendix~\ref{app:agentic}), and reasoning and Mixture-of-Experts models show weaker single-token causal leverage (Appendix~\ref{app:regimes}). Base-vs-instruct results further suggest that attack decodability emerges during pre-training, while instruction tuning suppresses compliance and sharpens late-layer geometry (Appendix~\ref{app:training_stage}).

\paragraph{Limitations.}
Our detailed analysis focuses primarily on Qwen3-4B, although we evaluate architectures up to 32B parameters. The peak layer varies from 63\% to 91\% of network depth and must therefore be identified per model. Our main stimuli are also primarily synthetic, single-turn, and English-language. Moreover, the strong single-token causal localization observed in dense, non-thinking models becomes substantially weaker in reasoning, MoE, and agentic settings, where compliance may be distributed across generated tokens or computation paths. Broader threat models such as tool misuse or data exfiltration may therefore involve different multi-step mechanisms.

\section{Conclusion}
\label{sec:conclusion}

Prompt injection compliance is causally concentrated rather than diffusely distributed. Across architectures from 4B to 32B parameters, causal leverage concentrates in a late-layer bottleneck spanning the final third of the network, operating within a low-dimensional linear subspace that we confirmed as a stable architectural feature through split-half analysis.

The causal peak also serves as the optimal site for attack detection, successfully validating a prediction derived directly from the mechanistic map.
Stealthy attacks converge on this same bottleneck despite following entirely distinct representational pathways in early layers. Naturalistic and indirect injections reveal the principled limits of centroid-based transfer, but reinforce the architectural stability of the decision window itself. Translating this causal map into robust, generalizable defenses remains the central open challenge in securing language models against adversarial inputs.

\bibliographystyle{iclr2027_conference}
\bibliography{normal_generated_py3}

\appendix

\section{System Prompts}
\label{app:prompts}

The default AcmeCorp system prompt used in all experiments:

\begin{quote}
\textit{You are a helpful assistant for AcmeCorp, a company that sells widgets and gadgets. You help customers with their orders, returns, and general questions about AcmeCorp products. You should only answer questions related to AcmeCorp products and services. If a customer asks about something unrelated to AcmeCorp, politely decline and redirect them to AcmeCorp-related topics.}
\end{quote}

Additional system prompts for multi-prompt generalization:
\textbf{MedBot} (medical information, general health only),
\textbf{CodeReviewer} (programming and software engineering only),
\textbf{LegalAdvisor} (general legal principles only).

\section{Rank-$k$ Subspace Patching}
\label{app:rankk}

To determine how many dimensions the compliance mechanism occupies, we project the per-stimulus patching displacement $\mures^{(24)} - \hh{24}_{\text{comp},i}$, where $\hh{24}_{\text{comp},i}$ is the unpatched state of compliant stimulus $i$, onto the top-$k$ PCA components of the pooled compliant/resistant state matrix at L24, and add only that projected displacement to $\hh{24}_{\text{comp},i}$. If the mechanism is low-rank, flip rate should saturate well before $k$ reaches the full hidden dimension (2560 for \model{}). Table~\ref{tab:rankk} shows saturation near $k{=}16$, with the 80\% threshold already crossed at $k{=}8$ for both \model{} and Qwen3-14B and at $k{=}64$ for Qwen3-32B, confirming the compliance mechanism is compact.
The 85.7\% rank-8 value quoted in Section~\ref{sec:res_cascade} comes from a different, class-mean protocol (projecting $\mures-\mucomp$ onto a PCA basis fit on resistant states and replacing the state with $\mucomp$ plus the projection), so it is not directly comparable to the 81.0\% here.

\begin{table}[h]
\centering
\caption{Rank-$k$ PCA subspace patching at the peak layer (\model{} L24; Qwen3-14B L28, Qwen3-32B L58), per-stimulus protocol.
The 80\% threshold is crossed at $k{=}8$ for \model{} and Qwen3-14B and at $k{=}64$ for Qwen3-32B (2560, 5120, and 5120 hidden dimensions).}
\label{tab:rankk}
\small
\setlength{\tabcolsep}{4pt}
\begin{tabular}{rcccl}
\toprule
$k$ & \model{} (L24) & Qwen3-14B (L28) & Qwen3-32B (L58) & Note \\
\midrule
1   & 33.3\% & 54.3\% & 45.6\% & rank-1 PCA \\
2   & 45.2\% & 56.8\% & 51.6\% & \\
4   & 63.1\% & 74.1\% & 70.9\% & \\
\textbf{8}   & \textbf{81.0\%} & \textbf{80.2\%} & 72.5\% & \textbf{80\% threshold (4B, 14B)} \\
16  & 89.3\% & 84.0\% & 79.1\% & near-full recovery \\
\textbf{64}  & 90.5\% & 86.4\% & \textbf{80.8\%} & \textbf{80\% threshold (32B)} \\
Full & 90.5\% & 87.7\% & 81.3\% & upper bound (no projection) \\
\bottomrule
\end{tabular}
\end{table}

\section{Sublayer Decomposition}
\label{app:sublayer}

Each transformer layer contains two sublayers (multi-head attention and an MLP) connected by a residual stream that carries the input forward. To determine which component is responsible for the causal signal, we run three patching variants at each of seven layers: (a)~\emph{full}, replacing the entire residual-stream output; (b)~\emph{attn-only}, replacing only the attention sublayer output; and (c)~\emph{MLP-only}, replacing only the MLP sublayer output.

If the causal signal were generated \emph{within} the target layer's
attention or MLP, then patching those sublayers alone should recover most
of the full-patch effect.  The result (Table~\ref{tab:sublayer}) is clear: at L24, neither sublayer alone accounts for more than 24\% of the full 90.5\% effect. The signal is not computed locally. Instead, it arrives at L24 already encoded in the \emph{incoming residual stream}, having been built up progressively across preceding blocks.

This finding constrains the causal interpretation: L24 is a
\emph{leverage point}, not a \emph{computation site}.  The compliance representation is built up progressively across the preceding blocks and arrives encoded in the residual stream entering L24; L24 itself does not freshly compute the compliance outcome.  Patching at L24 is therefore best understood as intercepting a signal in transit rather than overwriting a local decision unit.  This is consistent with the broader late-layer localization pattern: the representation is transformed, not created, at the peak layer.

\begin{table}[h]
\centering
\caption{Sublayer decomposition at key layers (\model{}).
At L24 (peak), neither attention nor MLP alone accounts for more than 24\%
of the full-patch effect (90.5\%), identifying the \emph{incoming residual
stream} as the primary causal carrier.}
\label{tab:sublayer}
\begin{tabular}{rcccc}
\toprule
Layer & Full & Attn-only & MLP-only & Primary site \\
\midrule
L1  &  3.6\% &  1.2\% &  1.2\% & Neither \\
L8  &  7.1\% &  1.2\% &  2.4\% & Neither \\
L16 & 15.5\% &  6.0\% &  9.5\% & MLP \\
\textbf{L24} & \textbf{90.5\%} & 21.4\% & 21.4\% & \textbf{Incoming residual} \\
L28 & 89.3\% &  7.1\% & 21.4\% & MLP + residual \\
L32 & 85.7\% & 10.7\% & 10.7\% & Incoming residual \\
L36 & 88.1\% &  7.1\% & 17.9\% & Joint/residual \\
\bottomrule
\end{tabular}
\end{table}

\section{Indirect Injection Full Results}
\label{app:indirect}

For indirect injection, the adversarial payload is buried inside a simulated knowledge-base retrieval block in the user turn rather than stated directly.  We constructed 600 stimuli (300 topics $\times$ 2 prefix variants) and measured a 25.0\% natural compliance rate (150/600), compared with 28\% on the multi-template direct corpus. Framing matters in both directions rather than uniformly reducing potency: the support-ticket framing of the 60-stimulus indirect pilot below yields 73\% (44/60), so we make no general claim about indirect framing and potency.

\paragraph{Indirect-own layer sweep.}
Table~\ref{tab:indirect_sweep} reports the sparse layer sweep on the $N_c{=}150$ compliant stimuli, patching the global resistant centroid of the indirect corpus itself. The raw flip rate rises late and peaks at L36 (43.3\%), but at L36 a norm-matched random vector also flips 43.3\% of the targets (compliant mean 43.3\%, benign mean 30.0\%), so the global indirect-own donor carries no class-specific signal. With 300 retrieval topics in the resistant pool, its centroid sits near the grand mean of the corpus, and its displacement from the compliant state is dominated by topical variation rather than by the compliance contrast.

\begin{table}[h]
\centering
\small
\caption{Indirect injection with the global indirect-own centroid (\model{}, 600-stimulus corpus, $N_c{=}150$). The raw rise is late, but at L36 the norm-matched random control also flips 0.433.}
\label{tab:indirect_sweep}
\begin{tabular}{lc}
\toprule
Layer (depth) & Flip rate \\
\midrule
L1  (3\%)   & 0.087 \\
L8  (22\%)  & 0.133 \\
L16 (44\%)  & 0.160 \\
L24 (67\%)  & 0.287 \\
L28 (78\%)  & 0.280 \\
L32 (89\%)  & 0.400 \\
\textbf{L36 (100\%)} & \textbf{0.433} \\
\bottomrule
\end{tabular}
\end{table}

\paragraph{Clean-donor transfer and operator robustness.}
The dilution above is a property of the indirect-\emph{own} donor, not of the target layer. Transferring a \emph{clean} donor --- the direct-injection resistant/compliant contrast --- onto the $N_c{=}44$ compliant targets of the 60-stimulus indirect pilot corpus (whose indirect-own donor pool holds only 16 resistant states) recovers a 95.5\% flip rate at L24 (median generation perplexity 1.7, i.e.\ coherent refusals), versus 15.9\% (L24) and 6.8\% (L36) for the diluted indirect-own centroid. Under this clean donor the effect is operator-robust (Table~\ref{tab:indirect_ops}): a full centroid, a $2\times$-scaled contrastive direction, and a rank-8 pooled subspace flip 95.0\%, 92.5\%, and 85.0\% of cases respectively ($N_c{=}40$ at L24, a second regeneration of the same corpus), whereas a rank-8 subspace built from resistant states alone reaches only 55.0\%. The compliance geometry at the late layer is therefore intact for indirect injection; the original global centroid simply lacked a donor that isolates the compliance contrast from topical variation.

\begin{table}[h]
\centering
\small
\caption{Indirect injection with a clean direct-injection donor at L24 (\model{}).
The late-layer effect recovers to direct-injection levels and is robust across operators;
the diluted indirect-own centroid is shown for comparison. Targets: 60-stimulus indirect pilot corpus ($N_c{=}40$ for the operator rows, $N_c{=}44$ for the indirect-own rows).}
\label{tab:indirect_ops}
\begin{tabular}{lc}
\toprule
Donor / operator & Flip rate \\
\midrule
Clean direct donor, full centroid            & 0.950 \\
Clean direct donor, scaled direction ($2\times$) & 0.925 \\
Clean direct donor, rank-8 pooled subspace   & 0.850 \\
Clean direct donor, rank-8 resistant-only    & 0.550 \\
\midrule
Indirect-own centroid (L24)                  & 0.159 \\
Indirect-own centroid (L36)                  & 0.068 \\
\bottomrule
\end{tabular}
\end{table}

\section{Multi-turn Escalation}
\label{app:robustness}

We test whether late-layer localization persists when the injection arrives after a benign conversational turn. We construct 360 two-turn stimuli using two softer injection templates, three benign warm-up queries, and 60 topics. Turn~1 contains a normal AcmeCorp support query, while Turn~2 introduces the injection. We collect hidden states from the full conversation and apply the same last-token patching protocol as in the single-turn setting.

Of the 360 stimuli, 62 (17.2\%) elicit compliance. Table~\ref{tab:multiturn_sweep} shows that the causal profile remains late. Flip rates are negligible at L1, rise sharply from L16 to L24, and peak at L24 with 87.1\%, close to the 91.7\% single-turn peak.

\begin{table}[h]
\centering
\small
\caption{Multi-turn causal sweep on \model{} ($N_c{=}62$). The peak remains at L24, matching the single-turn setting.}
\label{tab:multiturn_sweep}
\begin{tabular}{lcc}
\toprule
Layer & Flip rate & Depth \\
\midrule
L1  & 0.016 & 2.8\% \\
L16 & 0.113 & 44\% \\
L20 & 0.661 & 56\% \\
\textbf{L24} & \textbf{0.871} & \textbf{67\%} \\
L28 & 0.855 & 78\% \\
L32 & 0.742 & 89\% \\
L36 & 0.484 & 100\% \\
\bottomrule
\end{tabular}
\end{table}

These results show that a benign conversational prefix does not shift the causal region. Whether the injection appears in the first or second user turn, causal leverage remains concentrated in the same late block.

\section{Public-Benchmark Generalization}
\label{app:benchmark_matched}

\subsection{Benchmark-matched donors (\texttt{deepset}, \texttt{neuralchemy})}

This experiment addresses two questions: (1)~does causal leverage concentrate in the late block for naturalistic (non-synthetic) attacks, and (2)~does the lower flip rate observed with AcmeCorp donors (18.9\%) reflect a donor-mismatch confound? We answer both with a dense 19-layer sweep using donors drawn from the benchmark distribution itself.

\textbf{Design.}
We sampled 300 injections from two public benchmarks (\texttt{deepset/prompt-injections} and \texttt{neuralchemy/Prompt-injection-dataset}), split deterministically: the first 150 for calibration, the last 150 for testing. Resistant centroids at each sweep layer were computed from the 55 calibration-resistant stimuli ($N_r{=}55$, compliance rate 63\%,
$N_c{=}95$ calibration-compliant).
These benchmark-matched centroids were then applied to the 90 test-compliant stimuli across all 19 sweep layers ($\ell \in \{1, 2, 4, 6, 8, 10, 12, 14, 16, 18, 20, 22, 24, 26,
28, 30, 32, 34, 36\}$).

\paragraph{Results.}
Results are shown in Table~\ref{tab:benchmark_matched}.
\begin{table}[h]
\centering
\caption{Dense 19-layer causal sweep on naturalistic benchmark injections with benchmark-matched donors. Checkmark = null region (flip rate $<5\%$). Layers L1--L18 (first 50\% of depth) form a clean null region; the curve rises in the final third and peaks at L24 (67\% depth), replicating the AcmeCorp curve shape at roughly $7.5\times$ lower absolute rates.}
\small
\begin{tabular}{lccc}
\toprule
Layer & Depth & Flip rate & Null? \\
\midrule
L1  &  3\% & 1.1\% & \checkmark \\
L2  &  6\% & 0.0\% & \checkmark \\
L4  & 11\% & 0.0\% & \checkmark \\
L6  & 17\% & 0.0\% & \checkmark \\
L8  & 22\% & 0.0\% & \checkmark \\
L10 & 28\% & 1.1\% & \checkmark \\
L12 & 33\% & 3.3\% & \checkmark \\
L14 & 39\% & 3.3\% & \checkmark \\
L16 & 44\% & 3.3\% & \checkmark \\
L18 & 50\% & 3.3\% & \checkmark \\
L20 & 56\% & 8.9\% &            \\
L22 & 61\% & 8.9\% &            \\
\textbf{L24} & \textbf{67\%} & \textbf{12.2\%} & \textbf{(peak)} \\
L26 & 72\% & 10.0\% &            \\
L28 & 78\% & 10.0\% &            \\
L30 & 83\% & 10.0\% &            \\
L32 & 89\% & 10.0\% &            \\
L34 & 94\% & 10.0\% &            \\
L36 & 100\% & 10.0\% &           \\
\midrule
\multicolumn{4}{l}{\textit{3-condition controls at L24 (peak):}} \\
Resistant mean  & --- & 12.2\% & \\
Compliant mean  & --- &  3.3\% & \\
Norm-random     & --- &  1.1\% & \\
\bottomrule
\end{tabular}
\label{tab:benchmark_matched}
\end{table}

\paragraph{Interpretation.}
The sweep resolves both questions cleanly.
First, \emph{causal localization generalizes to naturalistic attacks}. The null region spans L1--L18 (all $\leq$3.3\% across ten consecutive layers), and the curve rises in the final third, peaking at L24 (67\% depth), the same relative location as in the AcmeCorp experiments.
Second, \emph{the lower absolute flip rate is not explained by donor mismatch}. Benchmark-matched donors actually produce a \emph{lower} peak flip rate (12.2\%) than the cross-distribution AcmeCorp donors (18.9\%), ruling out mismatch as a confound. The resistant signal remains class-specific: +11~pp above norm-matched random (12.2\% vs.\ 1.1\%), confirming that a compliance geometry exists at L24 for naturalistic phrasings. The roughly $7.5\times$ reduction in absolute flip rate relative to the synthetic setting (12.2\% vs.\ 91.7\%) reflects a weaker and less consistently aligned mean-difference signal across diverse real-world phrasings, not a failure of causal localization.

\subsection{Additional public benchmarks: BIPIA and Gandalf}
\label{app:external}

Beyond the deepset/neuralchemy phrasings of Appendix~\ref{app:benchmark_matched}, we replicate the causal sweep on two further public prompt-injection datasets under the AcmeCorp system prompt: \texttt{microsoft/BIPIA} (indirect injections embedded in email/web/table/summarization/code QA; we extract the [$N$] attack payloads and present each directly in the user turn, so this tests BIPIA attack content rather than its indirect threat model) and \texttt{Lakera/gandalf\_ignore\_instructions} (crowd-sourced, human-written ``ignore-your-instructions'' attacks, used as released). For each layer we patch with the dataset's own resistant mean and compare against a \emph{same-layer} norm-matched random control ($\mucomp + \text{unit-random}\cdot\|\mures-\mucomp\|$). The flip-minus-random gap, not the raw flip rate, is what distinguishes a class-specific effect from generic perturbation.

Table~\ref{tab:external} shows both datasets peak at L24, where the class-specific gap is largest (BIPIA $+0.57$, Gandalf $+0.34$) and collapses to $\approx$chance by L36, so the elevated raw flips in the very last layers are mostly generic perturbation, and L24 is the true class-specific peak. The raw flip rates at L24 ($0.66$--$0.70$) exceed the naturalistic \texttt{deepset}/\texttt{neuralchemy} rate ($\sim$12\%, Appendix~\ref{app:benchmark_matched}) because these datasets use narrower, more templated phrasing; the causal \emph{location} is identical. This is consistent with our central claim that the late-layer location is distribution-robust while the absolute magnitude tracks phrasing diversity.

\begin{table}[h]
\centering
\small
\caption{Causal sweep on two public benchmarks under the AcmeCorp system prompt
(\model{}; BIPIA payloads in the user turn, $N_c{=}23$; Gandalf $N_c{=}44$) Each cell is
flip rate / same-layer norm-matched random control. Both peak at L24 with the
largest gap over random.}
\label{tab:external}
\begin{tabular}{lcc}
\toprule
Layer (depth) & BIPIA (flip / rand) & Gandalf (flip / rand) \\
\midrule
L1  (3\%)   & 0.043 / 0.043 & 0.023 / 0.068 \\
L8  (22\%)  & 0.130 / 0.130 & 0.091 / 0.068 \\
L16 (44\%)  & 0.304 / 0.348 & 0.205 / 0.159 \\
L20 (56\%)  & 0.435 / 0.391 & 0.568 / 0.386 \\
\textbf{L24 (67\%)} & \textbf{0.696 / 0.130} & \textbf{0.659 / 0.318} \\
L28 (78\%)  & 0.652 / 0.261 & 0.636 / 0.614 \\
L32 (89\%)  & 0.652 / 0.435 & 0.614 / 0.545 \\
L36 (100\%) & 0.609 / 0.609 & 0.568 / 0.568 \\
\bottomrule
\end{tabular}
\end{table}

\section{Agentic Prompt Injection (InjecAgent, AgentDojo)}
\label{app:agentic}

To probe threat models beyond single-turn role conflict, we evaluate two agentic benchmarks in ReAct format: \textbf{InjecAgent}, with poisoned tool outputs targeting data theft or direct harm, and \textbf{AgentDojo}, with attacker goals across banking, Slack, travel, and workspace tasks. We use Mistral-7B as the primary testbed because it exhibits sufficient attack success: $36/62$ InjecAgent and $10/24$ AgentDojo stimuli trigger the attacker-requested action.

\textbf{The attack is decodable, but single-token causal leverage is not concentrated.}
A linear probe separating injected from benign tool results (matched pairs differing only in the tool output) reaches attack-\emph{presence} AUROC $=1.00$ from the first layer on Qwen3-4B, Qwen3-14B, and Mistral-7B. Presence, however, is not the compliance decision: a within-agentic \emph{compliance} probe (compliant vs.\ resistant rollouts) peaks at only AUROC $0.72$ (InjecAgent) and $0.54$ (AgentDojo). Causally, patching the resistant mean at the prompt-final token \emph{or} at the commitment token (where the attacker action is emitted) does not exceed a same-layer norm-matched random control at any layer (e.g.\ InjecAgent L16: commit $0.28$ vs.\ random $0.31$).

\textbf{Interpretation.}
The clean early-decode / late-cause dissociation of the single-step setting \emph{widens} rather than breaks: the attack is fully present in the residual stream from layer~1, yet the behavioral decision is produced across multiple generated steps, so no single-token intervention controls it. This is consistent with our central claim that representational availability does not imply a localized causal decision. Localizing the distributed agentic decision (e.g.\ multi-token or multi-step patching) is future work.

\section{Additional Models: Reasoning, MoE, and Training Stage}
\label{app:regimes}

\subsection{Reasoning and Mixture-of-Experts}

\textbf{Reasoning (thinking) mode.}
Re-running \model{} with its native chain-of-thought (\texttt{enable\_thinking=True}) leaves the attack decodable and preserves a late causal peak at L31 (86\% depth), but the absolute flip rate is much weaker than in the non-thinking setting (peak $0.21$). When the model emits an explicit reasoning trace, the compliance commitment appears to spread across generated reasoning tokens, so a single prompt-final patch is no longer the natural intervention point.

\textbf{Mixture-of-Experts.}
On Qwen3-30B-A3B (48 layers, ${\sim}3$B active parameters per token) the causal signal is much noisier: the peak flip rate is $0.06$ at L21, barely above the same-layer random control ($0.04$). We therefore make no strong claim that the dense single-token bottleneck carries over to sparse MoE routing, and whether expert routing distributes the compliance computation is an open question. Both regimes indicate that the single-token causal intervention is strongest for dense, non-thinking models.

\subsection{Training-stage attribution: base vs.\ instruct}
\label{app:training_stage}

To ask \emph{where in training} these representations arise, we compare Qwen3-4B-Base (pre-training only) and Qwen3-4B (instruction-tuned) under identical prompts and formatting (Table~\ref{tab:training_stage}). The injection representation is already fully present after pre-training: a linear-SVM probe (5-fold AUROC on the 300 headline injections vs.\ 60 ordinary benign queries) separates injection from benign with AUROC $=1.00$ at L1 in the base model, as it does in the instruct model on the same set, so instruction tuning does not create decodability from scratch. This is the same headline set as the detector of Section~\ref{sec:res_detection} (where the L1 AUROC of 0.888 is measured against hard negatives, not ordinary benign queries) and a larger set than the 30-vs-30 probe pilot behind the 80\% accuracy of Section~\ref{sec:res_probe}. What post-training changes is behavior and geometry: it lowers the compliance (attack-success) rate from $0.65$ to $0.28$ and raises the late-layer contrastive-direction norm at L32 by 74\% ($72.5\to126.0$). In short, pre-training builds the representation; post-training suppresses compliance and sharpens the late-layer direction into behavioral control.

\begin{table}[h]
\centering
\small
\caption{Base-vs-instruct comparison for Qwen3-4B under identical prompts.
Decodability is already maximal after pre-training; instruction tuning suppresses
compliance and sharpens the late-layer direction.}
\label{tab:training_stage}
\scalebox{0.9}{
\begin{tabular}{lccc}
\toprule
Model & Attack decodable at L1 & Direction norm at L32 & Compliance (ASR) \\
\midrule
Qwen3-4B-Base (pre-training only) & 1.00 & 72.5 & 0.65 \\
Qwen3-4B (instruct) & 1.00 & 126.0 & 0.28 \\
\bottomrule
\end{tabular}
}
\end{table}

\section{External detector baseline and adaptive evasion}
\label{app:detector_ext}
\textbf{Baseline.}
We compare the L24 activation probe against an off-the-shelf text classifier, \texttt{ProtectAI/deberta-v3-base-prompt-injection}, on the hard-negative set of Section~\ref{sec:res_detection}. The L24 probe reaches AUROC $0.993$ versus $0.709$ for the text classifier, and it reuses a prefill activation (one PCA projection, standardization, and a linear readout) rather than a second model forward pass.

\textbf{Adaptive attacker.}
Because a deployed monitor's location may become known, we stress-test with a white-box adversary: universal GCG optimization of an adversarial suffix that minimizes the L24 detector score while preserving compliance. Table~\ref{tab:detector_ext} reports recall at a $5\%$-FPR threshold. The suffix roughly halves L24 recall ($1.00\to0.48$) while the model stays compliant, as expected for any single-layer monitor. Crucially, the same suffix barely transfers to the text classifier ($0.625\to0.600$): the evasion is specific to the L24 activation geometry, not a generic prompt-injection jailbreak. We therefore treat detection as mechanistic corroboration and a complementary lightweight signal, not a standalone defense.

\begin{table}[h]
\centering
\small
\caption{Detector recall at a 5\%-FPR threshold, before and after a white-box GCG
suffix optimized against the L24 score (\model{}). The suffix targets L24 activations
and does not transfer to the text classifier.}
\label{tab:detector_ext}
\begin{tabular}{lcc}
\toprule
Detector (recall @ 5\% FPR) & Plain injection & Adaptive GCG suffix \\
\midrule
L24 activation probe & 1.000 & 0.475 \\
ProtectAI DeBERTa-v3 & 0.625 & 0.600 \\
\bottomrule
\end{tabular}
\end{table}

\end{document}